
\font\subtit=cmr12
\font\name=cmr8
\input harvmac
\def\LMU#1#2#3{\TITLE{LMU-TPW \number\yearltd-#1}{#2}{#3}}
\def\TITLE#1#2#3{\nopagenumbers\abstractfont\hsize=\hstitle\rightline{#1}%
\vskip 1in\centerline{\subtit #2}%
\vskip 1pt
\centerline{\subtit #3}\abstractfont\vskip .5in\pageno=0}%
\LMU{24} {MULTIVALUED FIELDS ON THE COMPLEX PLANE}
{AND BRAID GROUP STATISTICS}
\centerline{F. F{\name ERRARI}\foot{\name Work supported
by the Consiglio Nazionale Ricerche, P.le A. Moro 7, Roma, Italy}}\smallskip
\centerline{\it Sektion Physik der Universit\"at M\"unchen}\smallskip
\centerline{\it Theresienstr. 37, 8000 M\"unchen 2}\smallskip
\centerline{\it Fed. Rep. Germany}
\vskip 2cm
\centerline{ABSTRACT}
{\narrower
We study in this paper a theory of free anyons associated to free
conformal field theories defined on Riemann surfaces with a discrete and
nonabelian group of authomorphisms. The particles are exchanged
according to a nonabelian statistics, in which the $R-$matrix satisfy a
multiparametric generalization of the usual Yang$-$Baxter equations.}
\Date{October 1992}
\newsec { INTRODUCTION}
\vskip 1cm
In two recent papers, \ref\ffdtre{F. Ferrari, {\it Phys. Lett}
{\bf 277B} (1992), 423.} and
\ref\partone{F. Ferrari, {\it Multivalued Fields on the Complex Plane
and Conformal Field Theories}, Preprint LMU-TPW 92-13.}, we have
introduced a method for constructing conformal blocks with given
monodromy properties, provided the monodromy properties can be related
to those of an affine algebraic curve defined on the complex sphere
${\rm\bf CP}_1$. The method was applied, as a first step, to the case in
which the algebraic curves have a discrete group of symmetry $D_m$.
Surely applications to more complicated curves are possible, but we do
not have the proof that the method works in the most general case.
Moreover, it is difficult to relate our conformal blocks to a conformal
field theory at genus zero, but it is relatively simple to associate
them with the theory of the $b-c$ systems on algebraic curves with $D_m$
group of symmetry. Also in the simple case discussed, the conformal
blocks, builded in terms of free fields and twist fields
\ref\sjm{
M. Sato, T. Miwa and M. Jimbo. Holonomic quantum fields
(Kyoto U.P. Kyoto), part I; 14 (1978) p. 223; II: 15 (1979) p. 201;
III: 15 (1979) p. 577; IV: 15 (1979) p. 871; V; 16 (1980) p.
531.}, \ref\others{
L. Dixon, D. Friedan, E. Martinec, S. Shenker,
{\it Nucl. Phys.} {\bf B282} (1987), 13;
S. Hamidi, C. Vafa, {\it Nucl. Phys.} {\bf B279} (1987), 465;
J. J. Atick, A. Sen, {\it Nucl. Phys.} {\bf B286} (1987), 189;
L. Bonora, M. Matone, F. Toppan and K. Wu, {\it Phys. Lett.}
{\bf 224B} (1989) 115; Nucl. Phys. {\bf B334} (1990) 717;
E. Guadagnini, M. Martellini, M. Mintchev, {\it Jour. Math. Phys.}
{\bf 31} (1990), 1226.}
that are
explicitly known, exhibit nontrivial properties when two of the twist
fields are exchanged. The twist fields turn out to be anions satisfying
a nonabelian statistics. The purpose of this paper is to study the braid
group statistics
\ref\froa{
J. Fr\"ohlich, Statistics of fields, The Yang-Baxter equation
and the theory
of knots and links, in: Nonperturbative quantum field theory, eds. G.
t'Hooft et al. (Plenum, New York, 1988);
Proceedings of the Gibbs Symposium, Yale University, D. G. Caldi, G. D.
Mostow (eds.),  (1990).}
of the twist fields.
\newsec{ CONFORMAL FIELD THEORIES WITH NONABELIAN GROUP OF SYMMETRY }
\vskip 1cm
In order to fix the notation, we briefly review the conformal field
theories with nonabelian group of symmetries introduced in
\partone\ and refs. \ref\ffbos
{F. Ferrari, {\it Jour. Math. Phys.} {\bf 32} (8), (1991) 2186.},
\ffdtre.
We begin considering the case in which the
nonabelian group of symmetry is simply $D_m$.
Further generalizations are possible as we will discuss in the next
Section.\vskip 1pt
First of all let us introduce the $b-c$ systems of spin $\lambda$ ($\lambda\in
{\rm\bf Z}$) on an algebraic curve $\Sigma_g$:
\eqn\bcac{S=\int_{\Sigma_g}d^2zb\bar\partial c+{\rm c.c.}}
$\Sigma_g$ is defined by the vanishing of the Weierstrass polynomial
\eqn\curve{y^{2m}-2q(z)y^m+q^2(z)-p(z)=0}
where $q(z)$ and $p(z)$ are polynomials of degree $mr$ and $2r'$
respectively.
The point $z=\infty$ is reached performing the substitution
$z'=1/z$, so that the variables $z$ and $z'$ represent a system of
local coordinates describing the complex sphere ${\rm\bf CP}_1$.
The genus $g$ of the curve $\Sigma_g$ is given by the Riemann-Hurwitz
formula:
\eqn\genusa{2g-2=2m((m-1)r-2)+2mr'\qquad\qquad mr\ge r'}
\eqn\genusb{2g-2=2r'(m-1)+2mr'-4m\quad\qquad\qquad mr\le r'}
The solution of eq. \curve\
is a multivalued function (on ${\rm\bf CP}_1$)
$y^{(l)}(z)$. The index $l=0,\ldots,2m-1$ denotes the branches of $y(z)$,
which are defined in the following way:
\eqn\ypsilon{
\cases{y^{(l)}(z)=e^{{2\pi i l\over m}}\enskip\root m \of{q(z)+ \sqrt{p(z)}}
\qquad 0\le l\le m-1\cr
y^{(l)}(z)=e^{{2\pi i l\over m}}\enskip\root m \of{q(z)- \sqrt{p(z)}}
\qquad m\le l\le 2m-5\cr}}
where $\epsilon^m=1$. The fact that the only two coordinates on $\Sigma_g$
are $z$ and $y(z)$, allows us to treat the $b-c$ systems as a multivalued
field theory on the complex sphere.
The function $y(z)$ has two different kinds of branch points.
The branch points $\alpha_i$, $i=1,\ldots,N_\alpha={\rm max}(2mr,2r')$
are the projections on ${\rm\bf CP}_1$ of the zeros of $y(z)$.
The branch points $\beta_j$, $j=1,\ldots,N_\beta=2r'$ represent
instead the roots of the polynomial $p(z)$. For simplicity the integers
$r$ and $r'$ are chosen in such a way that $z=\infty$ is never
a branch point.
The local group of monodromy, describing the sequence in which the branches of
the curve $y(z)$ are exchanged at the branch points, contains the nonabelian
group $D_m$ as a subgroup.
\smallskip
If we put $\lambda>0$, a basis of meromorphic $\lambda$-differentials
and $1-\lambda$-differentials in which all the elements have an independent
behavior at the branch points, marked by the index $0\le k\le 2m-1$,
is given by \partone,\ffdtre:
\eqn\basis{\eqalign{B_k^{(l)}(z)dz^\lambda=&
\left(y^{(l)}(z)\right)^{mq_{k,\alpha_i}}\left(p(z)
\right)^{q_{k,\beta_j}}dz^\lambda\cr
C_k^{(l)}(z)dz^{1-\lambda}=&
\left(y^{(l)}(z)\right)^{-mq_{k,\alpha_i}}\left(p(z)
\right)^{-q_{k,\beta_j}}dz^{1-\lambda}\cr}}
where
\eqn\qak{q_{k,\alpha_i}={[k]_m+\lambda(1-m)\over m}\qquad\qquad\qquad
[k_m]=[k+m]_m=k}
and
\eqn\qbk{\eqalign{q_{k,\beta_j}=&-{\lambda\over 2}\qquad\qquad\qquad
k=0,\ldots,m-1
\cr
=&{1-\lambda\over 2}\qquad\qquad\qquad k=m,\ldots,2m-1\cr}}
In eq. \basis\ all the possible monodromy properties around
the branch points that a
meromorphic tensor can exhibit on $\Sigma_g$ are represented.
Moreover any meromorphic tensor on $\Sigma_g$
can be expanded in terms of the elements of the basis \basis, the
coefficients of this expansion being rational singlevalued functions of
$z$ \partone.
As a consequence the $n$-point functions of the $b-c$ systems on
an algebraic curve,
explicitly computed using the method of fermionic construction
\partone, \ref\ffstr{F. Ferrari, {\it Int. Jour. Mod. Phys.}
{\bf A5} (1990), 2799.},
are also decomposable in $n$-tensor products of the modes
$B_k^{(l)}(z)dz^\lambda$ and $C_k^{(l)}(z)dz^{1-\lambda}$.
In fact, the building blocks entering the $n$-point functions
in the case $\lambda>1$ are
the following tensors:
\eqn\keylambda{K^{(ll')}_\lambda(z,w)dz^\lambda dw^{1-\lambda}={1\over 2m}
{dz^\lambda dw^{1-\lambda}\over z-w}\sum\limits_{k=0}^{2m-1}
B_k^{(l)}(z)C_k^{(l')}(w)}
and the zero modes
\eqn\zeromodes{\Omega_{i_k,k}(z)dz^\lambda=z^{i-1}B_k(z)dz^\lambda\qquad
1\le i_k \le N_{b_k}}
$K_\lambda^{(ll')}(z,w)dz^\lambda dw^{1-\lambda}$ is characterized by a
single pole in the variable $z$ when $z=w$ and $l=l'$.
In eq. \zeromodes\ the index $i_k$ runs over all the zero modes
$\Omega_{i_k,k}(z)dz^\lambda$ sharing the same monodromy properties of
$B_k(z)dz^\lambda$. The number of these zero modes $N_{b_k}$ can be
explicitly calculated in two ways: by explicit construction of the zero
modes as in ref. \partone\ or from the residues of the ghost current
$J^{(l)}(z)=:b(z)c(z):$ as we will see below.
The case $\lambda=1$ can be treated in an analogous way provided we
replace $K_\lambda^{(ll')}(z,w)dz^\lambda dw^{1-\lambda}$ with the
differential of the third kind:
\eqn\thirdkind{\omega_{a_{(l')}b_{(l'')}}^{(l)}(z)dz=
K_{\lambda=1}^{(ll')}(z,a)dz-
K_{\lambda=1}^{(ll'')}(z,b)dz}
This differential has two simple poles in $z=a$ and $z=b$ when $l=l'$ and
$l=l''$ respectively. Moreover, when $\lambda=1$, one has to consider
also the zero mode in the $c-$fields, given by $C_0(z)={\rm
const.}dz^0$, so that $N_{c_k}=\delta_{k,0}\delta_{1-\lambda,0}$.
This decomposition in $2m$ $k-$sectors, corresponding to the different
elements of the basis
\basis, is also evident in
the vacuum expectation values (vev) of the ghost current and of the
energy momentum tensor $T(z)$. In fact, it is possible to pick up in
these vev's the contributions $<T_k^{(l)}(z)>$ and $<J_k^{(l)}(z)>$
coming from the modes $B_k(z)$
that correspond to a given monodromy behavior at the branch points:
\eqn\splitting{<T^{(l)}(z)>={1\over 2m}\sum\limits_{k=0}^{2m-1}
<T_k^{(l)}(z)>\qquad\qquad\qquad
<J^{(l)}(z)>={1\over 2m}\sum\limits_{k=0}^{2m-1}
<J_k^{(l)}(z)>}
Since it will become useful for later purposes, we give the explicit
expression
of the vev $<J_k^{(l)}(z)>$:
\eqn\jklz{<J_k^{(l)}(z)>dz={1\over 2m}
\partial_z{\rm log}C_k^{(l)}(z)dz}
The vev $<T_k^{(l)}(z)>$ has been already derived in \partone\ and we do
not report it here.
It is possible to consider $<T_k^{(l)}(z)>$ and $<J_k^{(l)}(z)>$ as the
vev's of the energy momentum and of the ghost current of the fields
belonging to a given $k-$sector, distinguished by their monodromy
behavior at the branch points.
For example the particle content of the $k-$th sector is
provided by the residues of $<J_k^{(l)}(z)>$.
In Section 5 of ref. \partone\ we showed that the total residue of
$<J_k^{(l)}(z)>$ at $z=\infty$ is $1-2\lambda$.
Moreover, in the case of a curve with nonabelian monodromy group, there
are also the residues $q_{k,\alpha_i}(l)$ and $q_{k,\beta_i}(l)$
at the branch points, which are dependent on the branch
$l$ of $J_k(z)$ in which they are computed:
\eqn\qakl{
\eqalign{q_{k,\alpha_i}(l)=&0\qquad\qquad\qquad0\le l\le m-1\cr
=& q_{k,\alpha_i}\qquad\qquad\qquad m\le l\le 2m-1\cr}}
and \eqn\qbkl{
q_{k,\beta_j}(l)=q_{k,\beta_j}\qquad\qquad\qquad 0\le l\le 2m-1}
The charges $q_{k,\alpha_i}$ and $q_{k,\beta_i}$ were already
given in eqs. \qak\ and \qbk.
Summing the residues of $<J_k^{(l)}(z)>$ over all the possible
branch points and over the branches $l=0,\ldots,2m-1$, we get
the following equation determining the numbers of zero modes $N_{b_k},
N_{c_k}$ in a given $k-$sector:
\eqn\residues{
N_{b_k}-N_{c_k}=1-2\lambda-\sum\limits_{l=0}^{2m-1}
\sum\limits_{i=1}^{N_\alpha}{1\over 2m}
q_{k,\alpha_i}(l)-\sum\limits_{j=1}^{N_\beta}q_{k,\beta_j}}
The term $1-2\lambda$ comes from the residue of the current $<J_k^{(l)}(z)>$
at $z=\infty$ and it is the usual residue one would expect
in the flat case.
The residues at the branch points, proportional to $q_{k,\alpha_i}(l)$ and
$q_{k,\beta_j}$, represent a nontrivial correction introduced
by the global topology of the Riemann surface $\Sigma_g$.
Finally, from eq. \residues, it is clear that
the total ghost charge in
a given $k-$sector requires the introduction of $N_{b_k}-N_{c_k}$
zero modes, otherwise the amplitudes of the $b-c$ systems vanish identically.
These $N_{b_k}+N_{c_k}$ zero modes are exactly those of eq. \zeromodes\
having the same behavior at the branch points
of a given element $B_k(z)dz^\lambda$ of the basis \basis.\smallskip
Finally, a study of the leading order poles of $<T_k(z)>$ indicates that
the effect of the branch points in the correlation functions
can be simulated in a way that we will show later
by the introduction of the socalled twist fields.
These fields are conformal operators depending on the branch points
and having fractional ghost charges. In the case in which the
monodromy group of the algebraic curve is nonabelian, the
dependence of the twist fields on the branch points becomes necessarily
nonlocal.\smallskip
One possibility to construct effectively a $b-c$ theory on $\Sigma_g$
in which the splitting of the fields is explicit,
consists in expanding the fields in
powers of $z$ and $y(z)$. This is the most general expansion allowed
on an algebraic curve.
Moreover, on an algebraic curve \curve\ there are only $2m$
rationally independent functions $f_k(z)$,
which are solution of a Riemann monodromy problem.
The function $y(z)$ and its powers should be linear combinations
of the $f_k(z)$'s,
the coefficients being rational functions of $z$ as explained above.
Expanding also the rational functions in powers of $z$ and collecting all
the terms in $f_k(z)$ together, $0\le k\le 2m-1$, we get the final form of the
fields:
\eqn\expb{\sum\limits_{k=0}^{2m-1}\sum\limits_{i=-\infty}^{\infty}
z^{-i-\lambda}f_k(z)b_{k,i}}
\eqn\expc{\sum\limits_{k=0}^{2m-1}\sum\limits_{i=-\infty}^{\infty}
z^{-i+\lambda-1}f_k(z)c_{k,i}}
Here $b_{k,i}$ and $c_{k,i}$ are arbitrary coefficients
that in the quantum
case become creation and annihilation operators of the fields propagating with
the same monodromy properties of $f_k(z)$.
This is the scheme used in \ref\ferope{F.Ferrari, {\it Operator
Formalism on Algebraic Curves}, Preprint MPI-Th/92-15.}
in order to construct an operator formalism for the $b-c$ systems on Riemann
surfaces. Still there is an infinite number of possible solutions
$f_k(z)$ of the Riemann monodromy problem, which are equivalent up to
rational functions of $z$.
The proper solutions for the $b-c$ systems are given in \basis.
Any other basis would produce an operator formalism in which the residues
\qakl\ and \qbkl\ of the currents $J_k(z)$ are shifted by integers.
However this would be in contradiction with eqs. \qakl, \qbkl\
obtained computing
explicitly the vev $<J_k(z)>$ with the method of the fermionic construction
\partone.
\newsec{BRAID GROUP STATISTICS OF THE $b-c$ SYSTEMS}
Another possibility of realizing the theory of the $b-c$ systems
on an algebraic curve in agreement with eq. \splitting\ and \residues\
is based on bosonization \ffbos,\ffdtre.
The details are described in ref. \partone.
See also \ref\knirev{V. G. Knizhnik, {\it Sov. Phys. Usp.} {\bf 32}(11)
(1989) 945.},
\ref\brzn{M. A. Bershadsky and A. O. Radul, {\it Int. Jour. Math.
Phys.} {\bf A2} (1987) 165.} for a study of the $Z_n$ symmetric curves.
Here we will mainly study the braid group statistics of the twist fields.
First of all we associate to each $k$-sector
with different monodromy properties
at the branch points the couple of free $b-c$ fields $b_k(z)dz^\lambda$
and $c_k(z)dz^{1-\lambda}$.
These fields take their values on the complex sphere and act on the vacua
$|0>_k$, $0\le k\le 2m-1$. The fields belonging to
different $k-$sectors do not interact. The presence of the branch points in
the amplitudes is simulated by the twist fields $V_k^{(l_i)}(\alpha_i)$ and
$V_k^{(l_j)}(\beta_j)$ for each sector $k$.
These are primary fields as it is shown in \ffdtre\ and \partone.
The two point function of the
fields $b_k$ and $c_k$ is defined as follows:
\eqn\gfk{
G_{\lambda,k}^{(ll')}(z,w)dz^\lambda dw^{1-\lambda}={B_k^{(l)}(z)
C_k^{(l')}(w)\over z-w}\prod\limits_{s=1}^{N_{b_k}}
{(z-z_{k,s})\over(w-z_{k,s})}\prod\limits_{s=1}^{N_{c_k}}{(w-w_{k,s})\over
(z-w_{k,s})}dz^\lambda dw^{1-\lambda}}
The meaning of eq. \gfk\ as a two point function is due to the fact that
this meromorphic tensor yields the
vev's of the currents $J_k(z)$ and of the energy momentum tensors $T_k(z)$
discussed in the previous Section (see also \partone).
Strictly speaking, however, this is not a true Green function on the
algebraic curve because the
structure of the poles is wrong. For example the pole in $z=w$ does not
occur only when $l=l'$ as it should be. However it is a Green function
on the complex plane for the multivalued sector $k$.
\vskip 1pt
It was shown in \partone\ that the explicit expression of eq. \gfk\ in terms
of twist fields and free $b-c$ fields is given by:
\eqn\gfktf{
G_{\lambda,k}^{(ll')}(z,w)dz^\lambda dw^{1-\lambda}={{}_k<0|b_k(z)c_k(w)
\prod\limits_{s=1}^{N_{b_k}}b(z_{k,s})
\prod\limits_{s'=1}^{N_{c_k}}c(z_{k,s'})
\prod\limits_{i=1}^{N_\alpha}V_k(\alpha_i)
\prod\limits_{j=1}^{N_\beta}V_k(\beta_j)|0>_k\over
{}_k<0|
\prod\limits_{s=1}^{N_{b_k}}b(z_{k,s})
\prod\limits_{s'=1}^{N_{c_k}}c(z_{k,s'})
\prod\limits_{i=1}^{N_\alpha}V_k(\alpha_i)
\prod\limits_{j=1}^{N_\beta}V_k(\beta_j)|0>_k}}
Bosonizing the fields $b_k(z)$ and $c_k(z)$ with the rules:
\eqn\bosoniz{b_k(z)=e^{-i\varphi_k(z)}\qquad\qquad c_k(z)=e^{i\varphi_k
(z)}\qquad\qquad <\varphi_k(z)\varphi_{k'}(z')>=-\delta_{kk'}{\rm log}
(z-z')}
we get the following form of the twist fields:
\eqn\vka{
V_k^{(l)}(\alpha_i)={\rm exp}\left[i\oint_{C_{\alpha_i}}dt
<J_k^{(l)}(t)>\varphi_k(t)\right]}
\eqn\vkb{
V_k^{(l)}(\beta_j)={\rm exp}\left[i\oint_{C_{\beta_j}}dt
<J_k^{(l)}(t)>\varphi_k(t)\right]=
e^{-iq_{k,\beta_j}\varphi_k(\beta_j)}}
The vev of $J_k^{(l)}(z)$ is given explicitly in eq. \jklz\ and
$C_{\alpha_i}$,
$C_{\beta_j}$ represent small paths surrounding the branch points
$\alpha_i$ and $\beta_j$ respectively.
The operators appearing in eqs. \vka\ and \vkb\ satisfy a nonstandard
statistics \partone:
\eqn\bgsglob{
V^{(l)}_k(\gamma)V^{(l')}_k(\gamma')=$$$$
{\rm exp}\left[-q_{k,\gamma}q_{k,\gamma'}\oint_{C_{\gamma}}ds
<J_k^{(l)}(s)>\oint_{C_{\gamma'}}ds'<J_k^{(l')}(s')>{\rm log}(s-s')
\right]V^{(l')}_k(\gamma')V^{(l)}_k(\gamma)}
where $\gamma$ and $\gamma'$ take their values on the set of branch points
$\{\alpha_i,\beta_j\}$.
In eq. \bgsglob, the dependence in the
statistics of the twist fields on the branch points is reminiscent
of the non-flat spacetime background on which the theory was originally
defined.
The $q$-parameter is a complicated convolution of three
distributions. This is not in contrast with ref. \froa,
where it was shown that
the dependence of the
$q-$parameter on the spatial coordinates is excluded in the flat
case due to the translational invariance of the local exchange relations.
First of all we are here on a Riemann surface with a nontrivial geometry.
Secondly, the algebra
\bgsglob, defined on a local patch of the Riemann surface which is
topologically
equivalent to a complex plane with punctures at the branch points
$\gamma$ and $\gamma'$, becomes:
\eqn\vgvg{
V^{(l)}_k(\gamma)V^{(l')}_k(\gamma')=
e^{i\pi q_{k,\gamma}(l)q_{k,\gamma'}
(l')}V^{(l')}_k(\gamma')V^{(l)}_k(\gamma)}
with $\gamma\ne \gamma'$ and $l\ne l'$.
To derive eq. \vgvg\ we have simply evaluated the two line
integrals occurring in eq. \bgsglob\
using the following two equations, that express the coefficients
appearing in eq. \curve\ in terms of the branch points:
\eqn\pz{
q^2(z)-p(z)=\prod\limits_{i=1}^{N_\alpha}(z-\alpha_i)\qquad\qquad\qquad
p(z)=\prod\limits_{i=j}^{N_\beta}(z-\beta_j)}
\eqn\qz{
q(z)=\left(\prod\limits_{i=1}^{N_\alpha}(z-\alpha_i)+
\prod\limits_{j=1}^{N_{\beta_j}}(z-\beta_j)\right)^{1\over 2}}
The remaining task is just a computation of residues.
As a result, we see that the local algebra \vgvg\ has a
$q-$parameter ${\rm exp}\left[ i\pi q_{k,\gamma}
(l)q_{k,\gamma'}(l')\right]$ in which the dependence on the spatial
coordinates is only appearing in the indices of the charges
$q_{k,\gamma}(l)$ and $q_{k,\gamma'}(l')$.
\smallskip
At this point it is convenient to forget for the moment the index $k$ of the
monodromy sector and to denote all the branch points with the symbol
$\gamma_i$, $1\le i\le N_\alpha+N_\beta$, in such a way that:
\eqn\posbp{
\eqalign{\gamma_i=&\alpha_i\qquad\qquad\qquad 1\le i\le N_\alpha\cr
=&\beta_{i-N_\alpha}\qquad\qquad\qquad N_\alpha+1\le i\le N_\alpha+
N_\beta\cr}}
Moreover we group the the two indices $l_i$ and $\gamma_i$,
describing the branch of
$V_k^{(l_i)}(\gamma_i)$ and the branch point $\gamma_i$
simulated by the twist field
respectively, in an unique index.
The new composed indices will be denoted using the capital latin
letters $I,J,K\ldots$.
In this condensed notation we have for example that
$V_k^{(l_i)}(\gamma_i)\equiv V^I$,
$V_k^{(l_j)}(\gamma_j)\equiv V^J$, $1\le i,j\le N_\alpha+N_\beta$.
Analogously $q_{k,\gamma_i}(l_i)\equiv q_I$ and so on.
A sum over the index $I$, containing the indices $l_i$ and $\gamma_i$,
$1\le i\le N_\alpha+N_\beta$,
should be understood as a sum over the two indices $i$ and $l_i$.
These two indices are totally independent. The subscript $i$ in $l_i$
is just a convenient notation when many twist fields are considered
altogether, otherwise we could set $l_i=a$, $l_j=b$ and so on,
$a,b=0,\ldots,2m-1$.
The compact form of eq. \vgvg\ becomes:
\eqn\vgvgcomp{V^IV^J=e^{i\pi q_Iq_J}V^JV^I\qquad\qquad I\ne J}
In eq. \vgvgcomp\ we have still not taken into account the radial
ordering (or time ordering) that it is understood in eq. \gfktf.
As usual, two branch points are said to be radial ordered, i.e.
$\gamma\prec \gamma'$, iff $|\gamma|<|\gamma'|$, the symbol $|\enskip|$
denoting the modulus of a complex number.
Using this definition of time-ordering and eq. \vgvgcomp\ we get:
\eqn\vgvgtime{
T\left( V^IV^J\right)=\left[e^{-i\pi q_Iq_J}
\theta(I-J)+
e^{+i\pi q_Iq_J}\theta(J-I)\right]T\left(V^JV^I\right)}
where $\theta(I-J)\equiv\theta(|\gamma_i|-|\gamma_j|)$
is the Heaviside theta function.
The algebra \vgvgtime\ is trivially associative despite of the fact that the
$q$-parameter ${\rm exp}(i\pi q_Iq_J)$ depends also on the indices $I$ and
$J$.
In fact one can easily see that the following equation holds
\eqn\associat{
T\left(V^IV^JV^M\right)={\rm
exp}\left[i\pi\left(\epsilon(I,J)q_Iq_J+\epsilon(I,M)q_Iq_M+
\epsilon(J,M)q_Jq_M\right)
\right]
T\left(V^MV^JV^I\right)}
independently of the way in which the twist fields in the lhs are permuted in
order to get the rhs.
In \associat\ we have defined:
\eqn\sig{\eqalign{\epsilon(I,J)=&+1\enskip {\rm if}\enskip
\gamma_i\prec \gamma_j\cr
=&-1 \enskip{\rm if}\enskip \gamma_j\prec\gamma_i\cr}}
The relations \vgvgtime\ resemble the equations defining the quantum
hyperplane of \ref\quantum{ Yu. I. Manin, {\it Comm. Math. Phys.}
{\bf 123} (1989), 163; J. Wess, B. Zumino, {\it Nucl. Phys.}
{\bf B} (Proc. Suppl.) {\bf 18B} (1990), 302;
W. Pusz, S. L. Woronowicz, {\it Rep. Math. Phys.} {\bf 27} (1989),
231.},
based on the quantum group $GL_q(N)$, where $q$ is a root of unity.
We can interpret eqs. \gfktf\ and \vgvgtime\ in the following way.
In the correlators of the $b-c$ systems on a Riemann surface with
nonabelian monodromy group the
dependence on the branch points becomes no longer local, but it is
expressed through the twist fields \vka\ and \vkb. These operators
replace the coordinates $\alpha_i$ and $\beta_j$ and can be considered
as {\it quantum coordinates}.
One is tempted therefore to treat the twist fields formally as
coordinates of a complex hyperplane and to
rewrite eq. \vgvgtime\ in a matricial form
following refs. \quantum:
\eqn\vgvgmatrix{T\left(V^IV^J\right)=\sum\limits_{M,N}Q_{MN}^{IJ}(q_I,q_J)
 T\left(V^MV^N\right)}
where
\eqn\qmatrix{
Q^{IJ}_{MN}(q_I,q_J)=
{\rm exp}(i\pi q_Iq_J)\hat R^{ij}_{mn}\left[
{\rm exp}(- i\pi q_{k,\gamma_i}(l_i)
q_{k,\gamma_j}(l_j))\right]\delta^{l_i}_{l_n}\delta^{l_j}_{l_m}}
Moreover $\hat R(q)$, with $q$ independent of the row indices $i,j$,
satisfies the quantum Yang-Baxter equation:
\eqn\yb{
\hat R_{12}\hat R_{23} \hat R_{12}=
\hat R_{23}\hat R_{12} \hat R_{23}}
and the sum over $M$ and $N$ in eq. \vgvgmatrix\ is intended as a sum
over the small indices $m$, $n$ and $l_m$, $l_n$.
The $\hat R_{mn}^{ij}$ component of $Q_{MN}^{IJ}$ acts on the indices
$i,j,m,n$ denoting the branch points $\gamma_i,\gamma_j,\gamma_m$ and
$\gamma_n$, while the permutation matrix
$\delta_{l_n}^{l_i}\delta_{l_m}^{l_j}$  acts on the branch indices
$l_i,l_j,l_m$ and $l_n$.
Explicitly eq. \vgvgmatrix\ reads:
$$T\left(V^{(l_i)}(\gamma_i)V^{(l_j)}(\gamma_j)\right)=
\sum\limits_{m,n=1}^{N_\alpha+N_\beta}
{\rm exp}(i\pi q_{k,\alpha_i}(l_i)q_{k,\beta_j}(l_j))$$
\eqn\vgvgexpl{
\hat R^{ij}_{mn}\left[
{\rm exp}(- i\pi q_{k,\gamma_i}(l_i)
q_{k,\gamma_j}(l_j))\right]\sum\limits_{i_m,i_n=0}^{2m-1}
\delta^{l_i}_{l_n}\delta^{l_j}_{l_m}
T\left(V^{(l_m)}(\gamma_m)V^{(l_n)}(\gamma_n)\right)}
where we remember that the indices
$l_m$ and $l_m$ are actually independent of $m$ and $n$.
We can also generalize eq. \vgvgexpl\ allowing for $q-$parameters which
are not roots of unity. In this case ${\rm exp}(i\pi q_I q_J)$ can be
substituted by any function $f(q_I,q_J)$ such that $f(q_I,q_J)=f(q_J,q_I)$
\smallskip
The only difference between eq. \vgvgmatrix\ and the quantum hyperplane
is that the $q-$parameter appearing in the matrix $Q_{MN}^{IJ}$
depends now also on the row indices $I$ and $J$.
The associativity property \associat\ requires that the matrix $Q$
of eq. \qmatrix\ still satisfies a relation similar to the Yang-Baxter
equations.
Indeed it is easy to prove that:
\eqn\ybgeneral{
Q_{12}(q_1,q_2)Q_{23}(q_2,q_3)Q_{12}(q_1,q_2)=
Q_{23}(q_2,q_3)Q_{12}(q_1,q_2)Q_{23}(q_2,q_3)}
This is not the usual Yang-Baxter equation, but a generalization of it that
was firstly found in ref. \ref\jimbo{M. Jimbo, {\it Int. Jour. Mod.
Phys.} {\bf A4} (1989), 3759.}.
In our case the form of eq. \ybgeneral\ is dictated by the fact that the
twist fields obey the nonabelian statistics \vgvgcomp, which is a consequence
of the nonabelian monodromy group of the algebraic curve \curve.
It is remarkable the fact that in refs.
\ref\mccoy{H. Au-Yang, B. M. Mc Coy, J. H. H. Perk,
S. Tang, M. L. Yan, {\it Phys. Lett.} {\bf A123} (1987), 219},
\ref\mccoytwo{R. J. Baxter, J. H. H. Perk, H. Au-Yang, {it Phys. Lett.}
{\bf A128} (1988), 138}. Explicit
solutions of eq. \ybgeneral\ were found from integrable models in which the
spectral parameter is defined on algebraic curves of genus $g>1$.
Multiparameter deformations of the quantum hyperplane such as in eq.
\vgvgtime\ were also discussed in \ref\zac{D. Fairlie, C. Zachos, {\it
Phys. Lett} {\bf 256B} (1991), 43; C. Zachos, Proceedings of the XX-th
International Conference on Differential Geometric Methods in
Theoretical Physics, Eds. S. Catto and A. Rocha, Vol. II, World
Scientific, (1992).} and \ref\sud{ A. Sudbery, {\it J. Phys.} {\bf A23}
(1990), L697.}.\smallskip
We conclude remembering that algebraic curves have many applications to
knot theory \ref\bknoer{ E. Brieskorn, H. Kn\"orrer, {\it Plane
Algebraic Curves}, Birkh\"auser Verlag (1986)}.
Here we have expressed in terms of twist fields
the braiding properties of the curve at the branch points.
The situation in which the curve becomes degenerate, for example when
two or more branch points coincide, is the most interesting for knot
theory. However this situation is difficult to realize using the twist
fields because the OPE's between two twist fields need some
regularization when both of them are concentrated at the same branch point.
\listrefs
\bye